\documentclass[aps,pra,reprint,superscriptaddress]{revtex4-1}

\usepackage{amsmath,amsthm,amsfonts,amssymb,bm}
\usepackage{times}
\usepackage[colorlinks={true},dvipdfm]{hyperref}
\hypersetup{citecolor={blue}, filecolor={blue}, linkcolor={blue}, urlcolor={blue}}
\usepackage{float}
\usepackage{epsf}
\usepackage{color}
\usepackage{verbatim}
\usepackage{enumerate}
\usepackage{multirow}
\usepackage{anysize}
\usepackage{graphicx}
\usepackage{comment}
\usepackage{bm}

\begin{document}

\title{Searching for quantum optimal controls in the presence of singular
critical points}
\author{Gregory Riviello}
\affiliation{Department of Chemistry, Princeton University, Princeton, New Jersey 08544, USA}
\author{Constantin Brif}
\affiliation{Department of Scalable \& Secure Systems Research, Sandia National Laboratories, Livermore, CA 94550, USA}
\author{Ruixing Long}
\affiliation{Department of Chemistry, Princeton University, Princeton, New Jersey 08544, USA}
\author{Re-Bing Wu}
\affiliation{Department of Automation, Tsinghua University and Center for Quantum Information Science and Technology, TNlist, Beijing, 100084, China}
\author{Katharine Moore Tibbetts}
\affiliation{Department of Chemistry, Princeton University, Princeton, New Jersey 08544, USA}
\affiliation{Department of Chemistry, Temple University, Philadelphia, PA 19122, USA}
\author{Tak-San Ho}
\affiliation{Department of Chemistry, Princeton University, Princeton, New Jersey 08544, USA}
\author{Herschel Rabitz}
\affiliation{Department of Chemistry, Princeton University, Princeton, New Jersey 08544, USA}

\begin{abstract}
Quantum optimal control has enjoyed wide success for a variety of theoretical and experimental objectives. These favorable results have been attributed to advantageous properties of the corresponding control landscapes, which are free from local optima if three conditions are met: (1) the quantum system is controllable, (2) the Jacobian of the map from the control field to the evolution operator is full rank, and (3) the control field is not constrained. This paper explores how gradient searches for globally optimal control fields are affected by deviations from assumption (2). In some quantum control problems, so-called singular critical points, at which the Jacobian is rank-deficient, may exist on the landscape. Using optimal control simulations, we show that search failure is only observed when a singular critical point is also a second-order trap, which occurs if the control problem meets additional conditions involving the system Hamiltonian and/or the control objective. All known second-order traps occur at constant control fields, and we also show that they only affect searches that originate very close to them. As a result, even when such traps exist on the control landscape, they are unlikely to affect well-designed gradient optimizations under realistic searching conditions.
\end{abstract}

\maketitle

\section{Introduction}
\label{sec:intro}

Over the last two decades, improvements to femtosecond lasers and pulse-shaping technology have led to exploration of the boundaries of quantum phenomena that can be effectively controlled in the laboratory \cite{Rabitz2000, LevisRabitz2002, Goswami2003, DantusLozovoy2004, BrixnerGerber2003, *Nuernberger2007, Brif2010NJP, Brif2012ACP, WollenhauptBaumert2011}. Successful optimal control experiments (OCEs) have been performed on a wide range of atomic, molecular, solid state, and biological targets, with applications including selective cleavage, rearrangement, and formation of chemical bonds \cite{Assion1998, *BergtBrixner1999, Levis2001, VajdaBartelt2001, *Daniel2003, Plenge2011, NuernbergerWolpert2010, *NuernbergerWolpert2012, MooreXing2013PCCP}, control of photoisomerization reactions \cite{VogtKrampert2005, *DietzekYartsev2006, *DietzekYartsev2007, ProkhorenkoNagy2006, *VogtNuernberger2006CPL, GreenfieldMcGrane2009}, coherent manipulation of soft X-rays produced via high-harmonic generation \cite{Bartels2000, *Bartels2001, *Bartels2004, Reitze2004, PfeiferKemmer2005, *PfeiferSpielmann2006, *Winterfeldt2008}, selective molecular electronic excitation \cite{BardeenYakovlevWilson1997, BrixnerDamrauer2001, Nahmias2005, Prokhorenko2005, KurodaKleiman2009, vanderWalleHerek2009, BonacinaWolf2007, RothGuyonRoslund2009, *RoslundRothGuyon2011, WeiseLindinger2011}, selective control of molecular vibrational states \cite{HornungMeierMotzkus2000, *WeinachtBartels2001CPL, *BartelsWeinacht2002PRL, KonradiSingh2006JPPA, *KonradiScaria2007, *ScariaKonradi2008, StrasfeldShim2007, *StrasfeldMiddleton2009}, control of energy flow in biomolecular complexes \cite{Herek2002, *WohllebenBuckup2003, *BuckupLebold2006, *SavolainenHerek2008}, and preservation of quantum coherence \cite{Branderhorst2008, BiercukBollinger2009}. Optimal control theory (OCT) \cite{Brif2010NJP, Brif2012ACP, RabitzZhu2000, WerschnikGross2007, BalintKurti2008, DAlessandro2007} provides a broad context to understand the experimental results and helps to gain insight into controlled dynamics of various quantum phenomena such as molecular photodissociation \cite{Kosloff1989, ShiRabitz1991, *Gross1991, *Gross1992, NakagamiOhtsuki2002, Krieger2011}, strong-field ionization \cite{RasanenMadsen2012}, photoisomerization \cite{OhtsukiOhara2003, ArtamonovHo2004CP, *ArtamonovHo2006CP, *ArtamonovHo2006JCP, *KurosakiArtamonov2009}, energy transport in light-harvesting complexes \cite{BruggemannMay2004CPL, *BruggemannMay2004JPCB, *BruggemannMay2006, *BruggemannMay2007, Caruso2012, HoyerWhaley2014}, electron ring currents in molecules \cite{KannoHoki2007}, photodesorption of molecules from a surface \cite{NakagamiOhtsukiFujimura2002CPL}, electron density transfer in one-electron \cite{Kammerlander2011} and many-electron systems \cite{Castro2012}, atom transport in optical lattices \cite{ChiaraCalarco2008, *DoriaCalarco2011, MischuckDeutsch2010}, manipulation of trapped Bose-Einstein condensates \cite{HohenesterRekdal2007, *HohenesterGrond2009, *GrondHohenester2009}, spin squeezing in atomic ensembles \cite{TrailDeutsch2010, *NorrisDeutsch2012}, and quantum information processing \cite{PalaoKosloff2002, *PalaoKosloff2003, TeschVivieRiedle2002, *VivieRiedleTroppmann2007, KhanejaReiss2005, *SchulteSporl2005, DominyRabitz2008JPA, Schirmer2009JMO, Nebendahl2009, *TsaiChenGoan2009, *ZhuFriedrich2013, Hohenester2006, GraceBrif2007JPB, *GraceBrif2007JMO, *GraceDominy2010NJP, *FloetherSchirmer2012, MontangeroCalarcoFazio2007PRL, *SafaeiMontangero2009, WeninPotz2008PRA, *WeninPotz2008PRB, *WeninRoloffPotz2009, *RoloffWeninPotz2009JCE, *RoloffWeninPotz2009JCTN, MerkelDeutsch2009, *DeutschJessen2010, *MischuckDeutsch2012, Rebentrost2009PRL, *RebentrostWilhelm2009, *MotzoiGambetta2009PRL, SchulteHerbruggenSporl2011, BrifGrace2013}. Also, optimal fields designed via OCT have been successfully used to control quantum information systems in experiments \cite{TimoneyElman2008, LuceroKelly2010, RowlandJones2012, LeeDeutsch2013, *SmithDeutsch2013, AtiaElias2013, ScheuerKong2014, ChoiDebnath2014}.

Both OCE and OCT are generally formalized as the search for a control field $\varepsilon(t)$ that produces the global maximum or minimum value of an \textit{objective functional} $J = J[\varepsilon(t)]$, which represents the control target. Typical quantum control objectives are the probability of a state-to-state transition, the expectation value of an observable, or the distance between the evolution operator and a target unitary transformation \cite{Brif2012ACP}. A number of recent studies \cite{RabitzHsiehRosenthal2004, HoRabitz2006JPPA, MooreHsiehRabitz2008JCP} strongly indicate that the success of numerous OCEs and OCT simulations is related to the favorable topology of the quantum control landscape defined by the functional dependence $J = J[\varepsilon(t)]$ \cite{Brif2012ACP, ChakrabartiRabitz2007review}. Experimental studies have illustrated the landscapes for various control problems \cite{VogtNuernberger2006PRA, *MarquetandNuernberger2007, RoslundRabitz2006, FormWhitaker2008, RoslundRabitz2009PRA_2, RuetzelStolzenberger2010, SchneiderWollenhaupt2011, MooreXing2013JCP}. The absence of local optima on the control landscape, which could trap a gradient search, is crucial for successful identification of a globally optimal control field. It has been shown \cite{HoRabitz2006JPPA, RabitzHoHsieh2006PRA, RabitzHsiehRosenthal2006JCP, *WuRabitzHsieh2008JPA, *HsiehWuRabitz2009JCP, RabitzHsiehRosenthal2005PRA, *HsiehRabitz2008PRA, *HoDominyRabitz2009PRA, *HsiehWuRabitzLidar2010} (also see Refs.~\cite{Brif2010NJP, Brif2012ACP, ChakrabartiRabitz2007review} for reviews) that quantum control landscapes for $N$-level closed systems are free from local traps if three conditions are satisfied: (1) the quantum system is \textit{controllable}, i.e., any unitary evolution operator can be produced in a finite time by application of admissible time-dependent controls; (2) the Jacobian matrix mapping the control field $\varepsilon(t)$ to the final-time evolution operator $U(T,0)$ is full rank everywhere on the landscape; (3) there are no constraints on the control field. We will discuss these conditions in more detail in Sec.~\ref{sec:back} below.

In this work, we assume that conditions (1) and (3) are satisfied and investigate how the violation of condition (2) affects the gradient-based optimization of various control objectives on closed, finite-level quantum systems. To this end, we perform large sets of numerical simulations that provide statistical evidence of the effect of singular critical points. While a number of recent studies considered control landscape topology \cite{Pechen2008JPA, *WuPechenRabitz2008JMP, *WuRabitz2012JPA, PechenBrif2010PRA} and optimization search effort \cite{OzaPechen2009JPA} for open quantum systems, issues concerning open-system control are beyond the scope of this work.

The remainder of this paper is organized as follows. Section~\ref{sec:back} contains background information on controllability and the characterization of landscape critical points. Section~\ref{sec:methods} describes the control objectives and computational methods employed in this work. In Sec.~\ref{sec:landscape}, we consider singular critical points that are characterized by a rank-deficient Jacobian violating condition (2), and explore their effect on trajectories of local gradient searches. Finally, our conclusions are summarized in Sec.~\ref{sec:concl}.

\section{Background}
\label{sec:back}

In this paper, we consider $N$-level closed quantum systems with Hamiltonians of the form
\begin{equation}
\label{eq:ham-1}
H(t) = H_0 + \sum_{i=1}^K \varepsilon_i (t) H_i ,
\end{equation}
where $H_0$ is the field-free Hamiltonian, the control fields $\{\varepsilon_i (t) \}_{i=1}^K$ are real-valued functions of time defined on the interval $[0,T]$, and $\{ H_i \}_{i=1}^K$ are Hermitian operators through which the control fields couple to the system. While the landscape analysis discussed in this section assumes that the Hamiltonian has this form, control problems in which the Hamiltonian contains a term that is quadratic in the control field have been successfully optimized using monotonically convergent algorithms \cite{BalintKurti2005, RenBalintKurti2006, LapertTehini2008, NakagamiMizumoto2008}. A general formulation of such algorithms for control problems with quadratic and higher-order field terms has been proposed \cite{HoRabitz2011}, and controllability criteria for such problems have been described \cite{Turinici2007}. These results suggest that the landscape topology for those control problems is amenable to optimal searches, although a formal landscape analysis has not been performed. 

In the Schr\"{o}dinger picture, the state of the system at a time $t$ is represented by the density matrix $\rho(t)$ or, for pure states, by the state vector $|\psi(t)\rangle$. The system evolution is given by $\rho(t) = U(t) \rho_0 U^{\dagger}(t)$, where $\rho_0 = \rho(0)$ is the initial density matrix and $U(t) \equiv U(t,0)$ is the time-evolution operator (also referred to as the propagator). For pure states, the evolution is given by $|\psi(t)\rangle = U(t) |\psi_0\rangle$, where $|\psi_0\rangle = |\psi(0)\rangle$ is the initial state vector. The evolution operator satisfies the Schr\"{o}dinger equation:
\begin{equation}
\label{eq:schro}
i \hbar \frac{d}{dt} U(t) = H(t) U(t) , \ \ \ U(0) = \mathbb{I} ,
\end{equation}
where $\mathbb{I}$ is the identity operator. A quantum system governed by Eq.~(\ref{eq:schro}) is called \textit{evolution-operator controllable} \cite{Brif2012ACP, DAlessandro2007} if for any unitary operator $W$ there exists a set of controls $\{\varepsilon_i (t) \}_{i=1}^K$ such that $W$ is the solution of the Schr\"{o}dinger equation (\ref{eq:schro}) at some finite time. For an $N$-level closed quantum system, a necessary and sufficient condition for evolution-operator controllability is that the Lie algebra generated by the set of operators $(i/\hbar)\{ H_0 , H_1, \ldots , H_K \}$ be u($N$) (or su($N$) for a traceless Hamiltonian) \cite{Ramakrishna1995, *SchirmerSolomon2002a, *SchirmerSolomon2002b, *Albertini2003, *Altafini2009}. The appearance of local traps on the control landscape due to the loss of controllability was recently studied \cite{WuHsiehRabitz2011PRA}; in the present work, we consider only controllable quantum systems. However, even when a system is in principle controllable, some states can be unreachable if control fields are constrained \cite{RivielloBrif-prep}, potentially preventing achievement of a global optimum. Thus, we also assume in this paper that no significant constraints are imposed on the control field except where specifically noted.

To simplify our notation, we consider one control field $\varepsilon(t)$ and one system-field coupling operator $\mu$ so that the Hamiltonian is of the form $H(t)  = H_0 - \mu \varepsilon(t)$. This form arises in the electric dipole approximation, and the generalization to several control fields is straightforward. \textit{Critical points} of a quantum control landscape are control fields at which the first-order functional derivative of the objective $J[\varepsilon(t)]$ with respect to the control field is zero for all time, i.e.,
\begin{equation}
\label{eq:crit}
\frac{\delta J}{\delta \varepsilon(t)} = 0 , \ \ \ \forall t \in [0,T] .
\end{equation}
Different types of critical points are characterized by properties of second- and higher-order functional derivatives of $J$ with respect to the control field.  They include global optima, local optima, and saddle points \cite{Brif2012ACP, ChakrabartiRabitz2007review}. This characterization of critical points determines the control landscape topology and is of practical importance, since local optima may trap gradient searches. In some situations, they can even hinder the convergence of global stochastic algorithms \cite{DigalakisMargaritis2001}. On the other hand, numerous OCT simulations have demonstrated that gradient searches can rapidly identify globally optimal solutions when the landscape is free from local optima \cite{MooreHsiehRabitz2008JCP, MooreChakrabarti2011, MooreRabitz2011}. Building upon these theoretical and numerical results, a gradient algorithm \cite{RoslundRabitz2009PRA} and a derandomized evolution strategy (a combination of stochastic and quasi-local search) \cite{RoslundShir2009PRA} were successfully adapted to OCEs and significantly improved the efficiency of laboratory quantum control.

Another important concept in control landscape topology is the classification of a critical point as regular or singular \cite{Brif2010NJP, Brif2012ACP, ChakrabartiRabitz2007review, BonnardChyba2003}. The objective $J$ can be represented as a function of the final-time evolution operator $U_T \equiv U(T)$, i.e., $J = J(U_T)$. In turn, the evolution operator $U_T$ is a functional of the control field: $U_T = U_T [\varepsilon(t)]$. Then, using the chain rule, the criticality condition of Eq.~\eqref{eq:crit} can be rewritten as:
\begin{equation}
\label{eq:crit-1}
\frac{\delta J}{\delta \varepsilon(t)}
= \left\langle \nabla J(U_T), \frac{\delta U_T}{\delta \varepsilon(t)} \right\rangle = 0 ,
 \ \ \ \forall t \in [0,T] ,
\end{equation}
where $\langle A , B \rangle = \mathrm{Tr} (A^{\dag} B)$ is the Hilbert-Schmidt inner product, $\nabla J(U_T)$ is the gradient of $J$ at $U_T$, and $\delta U_T / \delta \varepsilon(t)$ is the first-order functional derivative of $U_T$ with respect to the control field. A critical point of $J[\varepsilon(t)]$ is called \emph{regular} if the Jacobian $\delta U_T / \delta \varepsilon(t)$ is full rank, and is called \emph{singular} if the rank of $\delta U_T / \delta \varepsilon(t)$ is deficient. Since $\delta U_T / \delta \varepsilon(t)$ can be written as \cite{HoRabitz2006JPPA}
\begin{equation}
\label{eq:UT-deriv}
\frac{\delta U_T}{\delta \varepsilon(t)} = \frac{i}{\hbar} U_T U^{\dag}(t) \mu U(t) ,
\end{equation}
the full rank condition for $\delta U_T / \delta \varepsilon(t)$ requires that all $N^2$ elements of $\mu(t) = U^{\dag}(t) \mu U(t)$ (as functions of time) be linearly independent over $[0,T]$. This condition is generally satisfied because . The Jacobian full rank condition is equivalent to the statement that the map $\delta U_T / \delta \varepsilon(t)$ from $\delta \varepsilon(t)$ to $\delta U_T$
\begin{equation}
\label{eq:surj}
\delta U_T = \int_0^T \frac{\delta U_T}{\delta \varepsilon(t)} \delta \varepsilon(t) dt
\end{equation}
is surjective, i.e, that any arbitrary variation $U_T \to U_T + \delta U_T$ is produced by some corresponding control variation $\varepsilon(t) \to \varepsilon(t) + \delta \varepsilon(t)$. Thus, satisfaction of condition (2) establishes sufficient freedom to locally vary the final-time propagator $U_T$ via the control field $\varepsilon(t)$.  

In addition, a critical point is either called \textit{kinematic} if $\nabla J(U_T) = 0$ or \textit{non-kinematic} if $\nabla J(U_T) \neq 0$. By definition, all regular critical points are kinematic. Among singular critical points some are kinematic and some are non-kinematic \cite{WuLongDominy2012}. If the quantum system is controllable and there are no constraints on the control field (i.e., conditions (1) and (3) are satisfied), then none of the regular critical points on the control landscape are local optima \cite{RabitzHoHsieh2006PRA, RabitzHsiehRosenthal2006JCP, *WuRabitzHsieh2008JPA, *HsiehWuRabitz2009JCP, RabitzHsiehRosenthal2005PRA, *HsiehRabitz2008PRA, *HoDominyRabitz2009PRA, *HsiehWuRabitzLidar2010, Brif2010NJP, Brif2012ACP, ChakrabartiRabitz2007review}. For state-transition control, all regular critical points are either global maxima or global minima. For observable control and evolution-operator control, all regular critical points are saddle points except for the global maximum and global minimum. For singular critical points, no such characterization is currently available in the literature. While it is known \cite{ChakrabartiRabitz2007review} that singular critical points may exist on quantum control landscapes, vast empirical evidence suggests that their measure is much smaller than that of regular ones (the special case of time-optimal control, which we do not consider here, is an exception \cite{BonnardChyba2003, LapertZhangBraun2010}). Therefore, until recently, the characterization of singular critical points has received very little attention. 

One recent work \cite{WuLongDominy2012} characterized singular controls as well as the necessary and sufficient conditions for them to be critical points.  While no simple test has been developed to determine whether the control landscape corresponding to a particular Hamiltonian and objective will contain singular critical points, that work introduced an algorithm designed to numerically identify singular critical points and located them for various control problems. None were found to be local traps. Two other recent works \cite{FouquieresSchirmer2013, PechenTannor2011} studied singular critical points at constant control fields and suggested that at least some of them may cause gradient searches to fail. In particular, it was shown that, for specially tailored system Hamiltonian and target observable, a singular critical point at zero field is a second-order trap (i.e., the Hessian of the objective with respect to the control field, $\mathsf{H}(t,t') = \delta^2 J/ \delta \varepsilon(t) \delta \varepsilon(t')$, is negative semidefinite at this point) \cite{PechenTannor2011, FouquieresSchirmer2013}. While the analysis of higher-order functional derivatives of the objective showed that this critical point is not a true local maximum \cite{PechenTannor2012IJC}, a second-order trap can, in principle, attract some gradient searches and thus hinder the achievement of a globally optimal solution. Although traps found at singular critical points exist under unusual physical conditions, it is worthwhile to evaluate their effect on the convergence of a local gradient algorithm. Therefore, in this paper we perform numerical OCT simulations for several control problems proposed in Refs.~\cite{PechenTannor2011, FouquieresSchirmer2013} and identify parameters that affect the probability of a successful optimization.

Even when a quantum system satisfies conditions (1) and (2) and thus lacks fundamental traps on its control landscape, constraints on the control field $\varepsilon(t)$ can impede optimization. Control constraints are unavoidable; in OCEs with lasers, for example, the nature of the optical source determines the available bandwidth and the design of the spatial light modulator in the pulse shaper determines the number and range of independent control variables. The numerical implementation of OCT requires discretization of the time-evolution of the system, which also acts as a constraint on $\varepsilon(t)$. If constraints are sufficiently severe, they may generate artificial local optima that prevent searches from reaching global optima on the landscape. Several numerical studies have explored the effects of serious control constraints \cite{MooreRabitz2012,MooreBrif2012}, but we do not focus on this subject in the present work.

\section{Methodology}
\label{sec:methods}

\subsection{Quantum control objectives and corresponding landscape topology}
\label{sec:structure}

Consider a closed, controllable $N$-level quantum system whose evolution is governed by Eq.~\eqref{eq:schro}. The goal of optimal control simulations and experiments is to find a control field $\varepsilon(t)$ that corresponds to a global maximum (or minimum) of the objective functional $J[\varepsilon(t)]$. OCT simulations in this work address three common quantum control objectives:
\begin{enumerate}[(I)]
\item State-transition control: maximizing the probability of a transition between initial and final pure states $| i \rangle$ and $| f \rangle$ at time $T$:
\begin{equation}
\label{eq:pif}
J_P = | \langle f | U_T | i \rangle | ^2 .
\end{equation}
\item Observable control: maximizing the expectation value of a quantum observable $\theta$ (a Hermitian operator) at time $T$:
\begin{equation}
\label{eq:tro}
J_{\theta} = \langle \theta(T) \rangle = \mathrm{Tr} \left( U_T^\dagger \theta U_T \rho_0 \right) .
\end{equation}
\item Evolution-operator control: minimizing the distance between the unitary evolution operator $U_T$ and a target unitary transformation $W$. Depending on the selected distance measure and normalization, this objective has been defined in several ways \cite{MooreBrif2012}, such as a phase-dependent form:
\begin{equation}
\label{eq:w}
J_W = \frac{1}{2} - \frac{1}{2N} \Re\, \mathrm{Tr} \left( W^{\dagger} U_T \right)
\end{equation}
or a phase-independent form:
\begin{equation}
\label{eq:w2}
\tilde{J}_W = 1 - \frac{1}{N} \left| \mathrm{Tr} \left( W^{\dag} U \right) \right| .
\end{equation}
\end{enumerate}
Note that $J_P$ is a special case of $J_{\theta}$ for which $\rho_0$ and $\theta$ are projectors onto the states $| i \rangle$ and $| f \rangle$, respectively. We will consider $\rho_0$ and $\theta$ that are diagonal in the eigenbasis of $H_0$ (except when noted otherwise), which causes no loss of generality in the control landscape analysis \cite{WuRabitzHsieh2008JPA}.

For each objective, the landscape analysis can be performed in the dynamic or kinematic formulation. In the \textit{dynamic formulation}, the control landscape $J = J[\varepsilon(t)]$ is defined on the $L^2$ space of control fields. In the \textit{kinematic formulation}, the control landscape $J = J(U_T)$ is defined on the unitary group U$(N)$ and treats the matrix elements of $U_T$ as control variables. If the Jacobian $\delta U_T/\delta \varepsilon(t)$ is full-rank at a critical point satisfying $\delta J / \delta \varepsilon(t) = 0$ in the dynamic formulation, then, according to Eq.~\eqref{eq:crit-1}, the corresponding critical point satisfying $\nabla J(U_T) = 0$ must exist in the kinematic formulation. More specifically, there exist multiple (generally, an infinite number of) control fields satisfying $\delta J / \delta \varepsilon(t) = 0$, all of which produce the same evolution operator satisfying $\nabla J(U_T) = 0$. Furthermore, the Hessian spectrum evaluated at a regular critical point has the same number of positive and negative eigenvalues in the dynamic and kinematic formulations \cite{WuRabitzHsieh2008JPA}. Therefore, if all critical points are regular, then the control landscape topology is identical in the kinematic and dynamic formulations.

The landscape topology for quantum control objectives (I)--(III) has been analyzed under the assumption that conditions (1)--(3) are met \cite{Brif2012ACP}, so it is sufficient to characterize critical points in the kinematic formulation. For state-transition control, the landscape $J_P(U_T)$ has only two critical points corresponding to the global maximum ($J_P = 1$) and minimum ($J_P = 0$) \cite{RabitzHsiehRosenthal2004, RabitzHoHsieh2006PRA}. For observable control, the landscape $J_{\theta}(U_T)$ generally has critical points other than the global maximum and minimum, but the eigenvalue spectrum of the Hessian reveals that all intermediate critical submanifolds are saddles \cite{HoRabitz2006JPPA, WuRabitzHsieh2008JPA}. The critical values of $J_{\theta}$ are determined by the eigenvalue spectra of $\rho_0$ and $\theta$. When $\rho_0$ and $\theta$ are projectors onto pure states, the topology of the observable landscape matches the state-transition landscape described above.  When $\rho_0$ and $\theta$ are full-rank operators \cite{HoRabitz2006JPPA, WuRabitzHsieh2008JPA}, the observable landscape has $N!$ critical points. For evolution-operator control, the landscape $J_W(U_T)$ has $N+1$ critical points corresponding to the objective values $J_W = 0, 1/N, 2/N, ... ,1$. The global minimum and maximum correspond to the values $J_W = 0$ and $J_W = 1$, respectively, while all intermediate critical submanifolds are saddles \cite{HsiehRabitz2008PRA, HoDominyRabitz2009PRA}. As described above, multiple control fields correspond to each critical point in the kinematic formulation.

\subsection{The optimization procedure}
\label{sec:contproc}

The goal of the OCT simulations in this paper is to optimize the objectives described in Eqs.~\eqref{eq:pif} -- \eqref{eq:w}. There exist numerous optimization algorithms, including global and local methods, many of which have been employed in OCE and OCT \cite{Brif2010NJP, Brif2012ACP}. Global methods (e.g., genetic algorithms and evolutionary strategies) are designed to escape local optima by stochastically sampling a large volume of the search space. Local methods include strategies such as the simplex and gradient algorithms; the latter has proven particularly efficient in OCT simulations. Gradient algorithms are deterministic and ``myopic'', i.e., each step of the search trajectory depends entirely on the local geometry at one point on the landscape.  This makes them ideal for probing local landscape features such as the traps that may arise when condition (2) is violated. Therefore, the simulations in this work exclusively use gradient methods.

In the procedure employed below \cite{MooreChakrabarti2011, MooreRabitz2011, MooreRabitz2012, RivielloRabitz-prep}, each search trajectory is parameterized by the algorithmic index $s$, which labels the changes made to the control field during the optimization. We use the notation $\varepsilon(s,t)$, where the value $s = 0$ corresponds to the initial field $\varepsilon_0(t)$. Subsequent fields along the search trajectory are generated by solving the initial value problem
\begin{equation}
\label{eq:searchalg}
\frac{\partial \varepsilon(s,t)}{\partial s} = \gamma \frac{\delta J[\varepsilon(s,t)]}{\delta \varepsilon(s,t)} , 
\quad \varepsilon(0,t) = \varepsilon_0(t) ,
\end{equation}
where $\gamma$ is a positive (negative) constant for maximization (minimization) of $J$. In numerical simulations, we solve Eq.~\eqref{eq:searchalg} using \texttt{ode45}, a fourth-order Runge-Kutta integrator with a variable step size, incorporated in MATLAB \cite{matlab}. The \texttt{ode45} routine requires the input of an absolute error tolerance, $\tau$, prior to an optimization, and uses $\tau$ to determine an appropriate step size for each algorithmic iteration.  All simulations in this work use a value of $\tau = 10^{-8}$, unless otherwise stated, which generally ensures accurate solutions to Eq.~\eqref{eq:searchalg}.

The functional derivative $\delta J / \delta \varepsilon(s,t)$ that appears in Eq.~\eqref{eq:searchalg} is computed using the chain rule, in the same way as in Eq.~\eqref{eq:crit-1}. Then, using the relationship in Eq.~\eqref{eq:UT-deriv} for control objectives \eqref{eq:pif} -- \eqref{eq:w}, one obtains \cite{HoRabitz2006JPPA, RabitzHoHsieh2006PRA, HoDominyRabitz2009PRA, MooreRabitz2012}:
\begin{align}
& \frac{\delta J_P}{\delta \varepsilon(t)} 
= \frac{2}{\hbar} \Im\, \left[ \langle f | U_T | i \rangle \langle i | \mu(t) U_T^{\dag} | f \rangle \right] , 
\label{eq:-grad-Jp} \\
& \frac{\delta J_{\theta}}{\delta \varepsilon(t)} 
= \frac{2}{\hbar} \Im\, \mathrm{Tr} \left[ U_T^\dagger \theta U_T \rho_0 \mu(t) \right] , \label{eq:-grad-Jtheta} \\
& \frac{\delta J_W}{\delta \varepsilon(t)} 
= \frac{1}{2N \hbar} \Im\, \mathrm{Tr} \left[ W^{\dagger} U_T \mu(t) \right] \label{eq:-grad-JW} .
\end{align}
The optimization is stopped when the search trajectory arrives at a control field $\varepsilon(s_f,t)$ that yields an objective value satisfying $J \geq (J_{\max} - \eta)$ or $J \leq (J_{\min} + \eta)$ for maximization or minimization of $J$, respectively.  $J_{\max}$ is the objective value corresponding to the global maximum of the control landscape while $J_{\min}$ corresponds to the global minimum. The convergence threshold $\eta = 0.001 \cdot (J_{\max} - J_{\min})$ is used in this work unless otherwise stated. The optimization \textit{search effort} is quantified as the number of algorithmic iterations required for convergence.

In principle, the control field $\varepsilon(t)$ is a continuous square-integrable function defined on the interval $[0,T]$. However, numerical simulations generally use a discrete, piecewise-constant representation of the field. This discretization is a constraint that can potentially interfere with the search for optimal controls \cite{RivielloBrif-prep}. Also note that Eqs.~\eqref{eq:UT-deriv} -- \eqref{eq:-grad-JW} are exact in the limit of continuous control fields, but are approximations for piecewise-constant fields. In this work, $\varepsilon(t)$ is defined at $L$ evenly spaced intervals: $\varepsilon(t) = \{ \varepsilon_l | t \in (t_{l-1},t_l] \}_{l=1}^L$, where $t_l = l \Delta t$ and $\Delta t = T/L$. Equation~\eqref{eq:schro} is numerically integrated by calculating the evolution operator $U(t_l) \equiv U(t_l,0)$ as a product of incremental propagators:
\begin{subequations}
\label{eq:calcu}
\begin{align}
& U(t_l) = U(t_l,t_{l-1}) \cdots U(t_2,t_1) U(t_1,t_0) , \\
& U(t_l,t_{l-1}) = \exp \left[ -\frac{i}{\hbar} (H_0 - \mu \varepsilon_l) \Delta t \right] ,
\end{align}
\end{subequations}
where $U_T = U(t_L)$. The control variables are the real, independently-addressable field values $\{\varepsilon_l\}$ at the $L$ time intervals. The gradient algorithm generates an evolving field vector $\{\varepsilon_l (s)\}$ along the search trajectory by solving the discretized version of Eq.~\eqref{eq:searchalg}:
\begin{equation}
\label{eq:searchalg-discr-1}
\frac{\partial \varepsilon_l(s)}{\partial s} = \gamma \frac{\partial J}{\partial \varepsilon_l(s)} , 
\end{equation}
where elements of the gradient vector are given by $\partial J / \partial \varepsilon_l = \Delta t\, \delta J / \delta \varepsilon(t_l)$, and the search starts from a vector of initial field values, $\{\varepsilon_l(0)\}$. The field values $\{ \varepsilon_l(s) \}$ are allowed to vary freely and independently at each step of the optimization algorithm after the initialization (i.e., for $s > 0$). At the start of each optimization, the field is initialized in the parameterized form below:
\begin{subequations}
\label{eq:field-init-1}
\begin{align}
& \varepsilon_0(t) = A(t) \sum_{m=1}^M a_m \cos (\omega_m t) , \\
& A(t) = A_0 \exp \left[ -(t-T/2)^2 / (2 \zeta^2) \right],
\end{align}
\end{subequations}
where $A(t)$ is the Gaussian envelope function and the parameter $\zeta$ determines its width. In this paper, we used $\zeta = T / 10$ (to ensure that $\varepsilon_0(t) \approx 0$ at $t = 0$ and $t = T$) and $M = 20$. The frequencies $\{ \omega_m \}$ were randomly selected from a uniform distribution on $[\omega_{\min},\omega_{\max}]$ (with $\omega_{\min}$ and $\omega_{\max}$ being the smallest and largest transition frequencies in $H_0$, respectively), and the amplitudes $\{ a_m \}$ were randomly selected from a uniform distribution on $[0, 1]$. The normalization constant $A_0$ was chosen so that the relative field strength (RFS, defined in Eq.~\eqref{eq:sigma} below) of each initial field is equal to a pre-selected value. All simulations use atomic units and Hamiltonians of the form $H(t)  = H_0 - \mu \varepsilon(t)$.

\section{Effect of singular critical points}
\label{sec:landscape}

Thousands of successful OCT runs have been performed for the objectives $J_W$ \cite{MooreChakrabarti2011}, $J_P$ \cite{MooreRabitz2011}, and $J_{\theta}$ \cite{RivielloRabitz-prep} on various controllable systems. Each optimization in these works used the gradient algorithm described in Sec.~\ref{sec:contproc} and converged successfully unless condition (3) was violated by significantly constraining the control field. These results suggest that control landscapes corresponding to the objectives above are generally free of local traps and thus amenable to a gradient search under proper conditions. However, there may be singular critical points (at which condition (2) is violated) on the landscape for various control problems \cite{WuLongDominy2012}.  The effect of these singular points on gradient-based searches is of interest in both OCE and OCT. 

This section explores how optimization convergence is affected by the presence of singular critical points. In particular, we perform sets of numerical searches on landscapes containing singular critical points that have been identified via Hessian spectral analysis \cite{PechenTannor2011, FouquieresSchirmer2013, PechenTannor2012IJC} as potential traps, in order to statistically evaluate the likelihood that they will prevent successful gradient optimization. We also perform a similar study for landscapes containing singular critical points whose Hessian spectra were not studied analytically, in order to determine whether they have any trapping effect on gradient searches.

\subsection{Singular critical points shown to be second-order traps}
\label{sec:traps}

Theoretical studies show \cite{PechenTannor2011, FouquieresSchirmer2013} that, for some control problems with specially tailored Hamiltonian and control objective, the landscape contains a singular critical point at zero field. At this point, the Hessian $\mathsf{H}(t,t')$ is negative (positive) semidefinite for an optimization goal of maximizing (minimizing) $J$. This landscape feature, whether it is only a second-order trap or a true local optimum, can potentially trap gradient searches. However, it has been argued that these artificially designed traps are unlikely to prevent optimization under realistic searching conditions \cite{PechenTannor2011Comm}. Here, we examine this issue in detail with five control problems whose landscapes have been proven to contain such a trap~\cite{PechenTannor2011, FouquieresSchirmer2013, PechenTannor2012IJC}. These problems are described below.
\begin{enumerate}[(A)]
\item Maximize $J_{\theta}$ (see Eq.~\eqref{eq:tro}) for a three-level $\Lambda$-type system:
\begin{equation*}
H_0 = \begin{pmatrix} E_1 & 0 & 0 \\ 0 & E_2 & 0 \\ 0 & 0 & E_3 \end{pmatrix}, \ 
\mu = \begin{pmatrix} 0 & 0 & \mu_{13} \\ 0 & 0 & \mu_{23} \\ \mu_{13} & \mu_{23} & 0 \end{pmatrix}.
\end{equation*}
For this problem, the landscape $J_{\theta}[\varepsilon(t)]$ has a second-order trap at $\varepsilon(t) = 0$ when the initial state and target observable are selected as $\rho_0 = |1 \rangle \langle 1 |$ and $\theta = \sum_{j=1}^3 \theta_j | j \rangle \langle j |$ with $\theta_2 > \theta_1 > \theta_3$, respectively \cite{PechenTannor2011}. The trap corresponds to the objective value $J = \theta_1$. In our simulations, we used values $E_1 = 0$, $E_2 = 10$, $E_3 = 30$, $\mu_{13} = 0.5$, $\mu_{23} = 1$, $\theta_1 = 0.3$, $\theta_2 = 0.5$, $\theta_3 = 0.2$. The final time is $T = 8$ and the time mesh is discretized into $L = 255$ intervals.

\item The same problem as (A), but with different parameter values \cite{PechenTannor2012IJC}: $E_1 = 0$, $E_2 = 1$, $E_3 = 2.5$, $\mu_{13} = -1$, $\mu_{23} = -1.7$, $\theta_1 = 0$, $\theta_2 = 1$, $\theta_3 = -5$, $T = 10$, and $L = 200$.

\item Maximize $J_P$ [see Eq.~\eqref{eq:pif}] for a four-level ladder-type system:
\begin{equation*}
H_0 = \begin{pmatrix} 2 & 0 & 0 & 0 \\ 0 & 4 & 0 & 0 \\ 0 & 0 & 5 & 0 \\ 0 & 0 & 0 & 9 \end{pmatrix}, \ 
\mu = \begin{pmatrix} 0 & -1 & 0 & 0 \\ -1 & 0 & -1 & 0 \\ 0 & -1 & 0 & -1 \\ 0 & 0 & -1 & 0 \end{pmatrix},
\end{equation*}
with initial and target states 
\begin{align*}
& | i \rangle = (\cos \varphi, 0, 0, \sin \varphi)^{\mathsf{T}} , \\
& | f \rangle = (e^{-2iT} \cos \vartheta, 0, 0, e^{-9iT} \sin \vartheta)^{\mathsf{T}} .
\end{align*}
For this problem, the landscape $J_P[\varepsilon(t)]$ has a second-order trap at $\varepsilon(t) = 0$ when $\vartheta$ and $\varphi$ lie in the interior of the same quadrant \cite{FouquieresSchirmer2013}. The trap corresponds to the objective value $J = \cos^2 (\vartheta - \varphi)$. In our simulations, we used $\vartheta = 1.58$, $\varphi = 3.08$, $T = 50$, and $L = 255$.

\item Maximize $J_P$ for another four-level ladder-type system
\begin{equation*}
H_0 = \begin{pmatrix} 1+\alpha & 0 & 0 & 0 \\ 0 & 1 & 0 & 0 \\ 0 & 0 & 2 & 0 \\ 0 & 0 & 0 & 2 \end{pmatrix}, \  
\mu = \begin{pmatrix} 0 & -1 & 0 & 0 \\ -1 & 0 & -1 & 0 \\ 0 & -1 & 0 & -b \\ 0 & 0 & -b & 0 \end{pmatrix},
\end{equation*}
with initial and target states
\begin{align*}
& | i \rangle = \frac{1}{\sqrt{2}} (e^{i \vartheta}, 0, 0, e^{-i \vartheta})^{\mathsf{T}} , \\
& | f \rangle = \frac{1}{\sqrt{2}} (e^{-i(1+\alpha)T}, 0, 0, e^{-2iT})^{\mathsf{T}} . 
\end{align*}
For this problem, the landscape $J_P[\varepsilon(t)]$ has a second-order trap at $\varepsilon(t) = 0$ when $T = \pi/\alpha$ and the inequalities $b^2 \cos^2(\vartheta) > (2/\pi) \sin(2 \vartheta) > 0$ are satisfied \cite{FouquieresSchirmer2013}. The trap corresponds to the objective value $J = \cos^2 \vartheta$. In our simulations, we used $\alpha = \pi/100$, $b = 3$, $\vartheta = \pi/3$, $T = 100$, and $L = 127$. Unlike for problems (A)--(C), in this problem the final time $T$ is fixed at half the period of the frequency $\alpha$. Since $\alpha$ corresponds to the $| 1 \rangle \leftrightarrow | 2 \rangle$ transition, fixing $T$ in this fashion makes it problematic to resolve the near degeneracy of levels $| 1 \rangle$ and $| 2 \rangle$, thereby placing a significant constraint on the control field and violating condition (3) for a trap-free landscape.  Such constraints have been shown to prevent successful optimization \cite{MooreBrif2012, RivielloBrif-prep}.

\item Minimize $J_W$ (see Eq.~\eqref{eq:w}) for a three-level ladder-type system with dynamic Stark shifts:
\begin{equation*}
H_0 = \begin{pmatrix} 1+\alpha & 0 & 0 \\ 0 & 1 & 0 \\ 0 & 0 & 2 \end{pmatrix}, \ 
\mu = \begin{pmatrix} -a & -1 & 0 \\ -1 & -b & -1 \\ 0 & -1 & -c \end{pmatrix}, 
\end{equation*}
with the target unitary transformation
\begin{equation*}
W = \begin{pmatrix} e^{-i \vartheta} & 0 & 0 \\ 0 & -ie^{-i \varphi} & 0 \\ 0 & 0 & -ie^{i \varphi} \end{pmatrix} e^{-i H_0 T/\hbar}.
\end{equation*}
For this problem, the landscape $J_W[\varepsilon(t)]$ has a second-order trap at $\varepsilon(t) = 0$ when $T = \pi/\alpha$ (just as in problem (D), defining $T$ in this way violates condition (3) by significantly constraining the control field) and when five coupled conditions involving $a$, $b$, $c$, $\vartheta$, and $\varphi$ are met \cite{FouquieresSchirmer2013, ZahedinejadSchirmer2014}. One combination that generates a trap is $a = 5 \sqrt{2/3}$, $b = 4$, $c = 1$, $\vartheta = 2 \pi/3$, and $\varphi = -3 \pi/4$; with this choice, the trap corresponds to the objective value $J = 5/12$. In our simulations, we used these values as well as $T = 1000$, $\alpha = \pi/1000$, and $L = 511$.
\end{enumerate}

While control problems (A) - (E) only contain a trap at $\varepsilon(t) = 0$, the local landscape geometry around this point on the landscape may prevent gradient optimizations from converging. In this work, a search was deemed to have failed if the monotonic improvement of the fidelity was interrupted, i.e., if the value of the objective functional decreased after an iteration. Defined in this way, \textit{search failure} indicates that the optimization method cannot solve Eq.~\eqref{eq:searchalg} with sufficient accuracy. A large body of successful OCT simulations \cite{MooreChakrabarti2011, MooreRabitz2011, RivielloRabitz-prep} shows that well-designed gradient searches rarely fail, but the presence of a second-order trap on the landscape may make search failure more likely.

The simulations reported below explore the effect of the traps in control problems (A)--(E) on gradient searches; the primary goal is to determine the relationship between the strength of the initial control field and the likelihood that a search beginning near the second-order trap will optimize. In order to compare fields across different control problems, we define the dimensionless \textit{relative field strength} (RFS):
\begin{equation}
\label{eq:sigma}
\sigma = \frac{1}{N_{\mu}} \sum_{\substack{i < j \\ \mu_{ij} \neq 0}} \frac{ \overline{|\varepsilon|} \times |\mu_{ij}|
}{|E_i - E_j|} ,
\end{equation} 
where the sum is over transitions with non-zero dipole matrix elements, $N_{\mu}$ is the number of these transitions, and $\overline{|\varepsilon|}$ is the mean field amplitude:
\begin{equation}
\label{eq:mfa}
\overline{|\varepsilon|} = \frac{1}{T} \| \varepsilon \|_1 = \frac{1}{T} \int_0^T |\varepsilon(t)| d t .
\end{equation}
In the discrete-time representation, $\overline{|\varepsilon|} = (1/L) \| \varepsilon \|_1 = (1/L) \sum_{l=1}^L |\varepsilon_l|$. The RFS is therefore the average ratio of the Rabi frequency to the transition frequency for the set of dipole transitions, and it is a natural metric of distance from a given field $\varepsilon(t)$ to the trap at zero field in control problems (A)--(E).

For each of the five control problems, we performed 10,000 optimization runs beginning at different initial fields $\varepsilon_0(t)$ with the same initial RFS $\sigma_0 = 1$. Table~\ref{tab:rfs1} reports the number of trapped searches and the mean search effort (MSE) for the 10,000 runs. Every search optimized successfully, despite the presence of a trap on the control landscape; these results suggest that initial fields with $\sigma_0 = 1$ are distant enough from the singular critical point $\varepsilon(t) = 0$ that the corresponding optimal searches succeed. To determine whether optimizations that begin closer to the trap may fail, additional searches were performed for $\sigma_0 \leq 1$ for control problems (A)--(E). 100 optimization runs were performed for each $\sigma_0$ value, and the results are reported in Table~\ref{tab:rfsvar}. The MSE and the average RFS of the optimized fields, $\overline{\sigma_{\text{opt}}}$, were recorded for all searches that successfully optimized.

\begin{table}[htbp]
\caption{\label{tab:rfs1}Optimization results for control problems whose landscape has a second-order trap at zero field. 10,000 optimization runs were performed for each control problem. Each search started from a random initial field of the form \eqref{eq:field-init-1}, with the RFS value $\sigma_0 = 1$.}
\begin{ruledtabular}
\begin{tabular}{ccr}
Problem & \# failed & MSE \\ \hline
(A) & 0 & 51 \\ 
(B) & 0 & 147 \\ 
(C) & 0 & 36 \\ 
(D) & 0 & 145 \\ 
(E) & 0 & 1492 \\ 
\end{tabular}
\end{ruledtabular}
\end{table}

\begin{table}[htbp]
\caption{\label{tab:rfsvar}Optimization results for various initial RFS values. 100 optimization runs were performed for each $\sigma_0$ value.}
\begin{ruledtabular}
\begin{tabular}{crrrr}
Problem & $\sigma_0$ & $\overline{\sigma_{\text{opt}}}$ & \# failed & MSE \\ \hline
\multirow{3}{*}{(A)} 
 & $5 \times 10^{-4}$ & $2.22 \times 10^{-2}$ & 0 & 1922 \\ 
 & $10^{-4}$ & $2.22 \times 10^{-2}$ & 98 & 11520 \\ 
 & $5 \times 10^{-5}$ & - & 100 & - \\ \hline
\multirow{3}{*}{(B)} 
 & $5 \times 10^{-3}$ & $2.12 \times 10^{-1}$ & 0 & 1697 \\ 
 & $10^{-3}$ & $2.14 \times 10^{-1}$ & 22 & 27485 \\ 
 & $5 \times 10^{-4}$ & - & 100 & - \\ \hline
\multirow{3}{*}{(C)}
 & $5 \times 10^{-5}$ & $5.45 \times 10^{-2}$ & 0 & 362 \\ 
 & $10^{-5}$ & $5.44 \times 10^{-2}$ & 53 & 1027 \\ 
 & $5 \times 10^{-6}$ & - & 100 & - \\ \hline
\multirow{3}{*}{(D)}
 & $5 \times 10^{-1}$ & $8.82 \times 10^{-1}$ & 0 & 137 \\ 
 & $10^{-1}$ & $6.91 \times 10^{-1}$ & 57 & 432 \\ 
 & $5 \times 10^{-2}$ & $6.80 \times 10^{-1}$ & 99 & 726 \\
 & $10^{-2}$ & - & 100 & - \\ \hline 
\multirow{4}{*}{(E)}
 & $5 \times 10^{-2}$ & $1.15$ & 0 & 4708 \\ 
 & $10^{-2}$ & $1.03$ & 86 & 2928 \\ 
 & $5 \times 10^{-3}$ & $1.01$ & 99 & 2725 \\ 
 & $10^{-3}$ & - & 100 & - \\
\end{tabular}
\end{ruledtabular}
\end{table}

\begin{figure}[htbp]
\centering
\includegraphics[width=8cm]{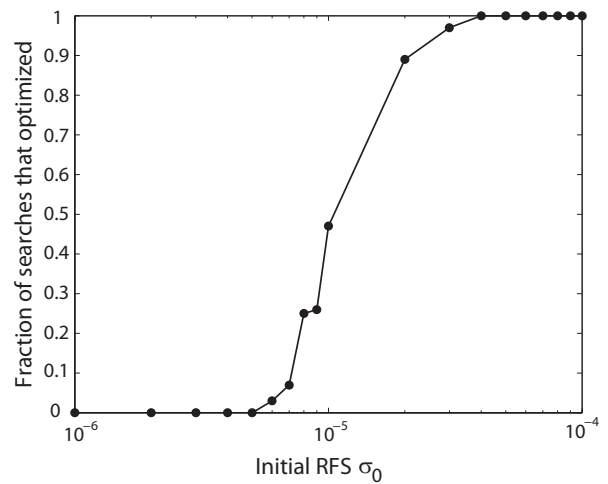}
\caption{The fraction of searches for control problem (C) that optimized successfully, as a function of the initial RFS $\sigma_0$. 100 optimization runs were performed for each value of $\sigma_0$.}
\label{fig:trapstrength}
\end{figure}
The simulations confirm that gradient optimizations of problems (A)--(E) with weaker initial fields (i.e., closer to the trap at $\varepsilon(t) = 0$) are increasingly likely to fail. More specifically, there exist RFS values $\sigma'$ and $\sigma''$ such that \emph{no} searches with $\sigma_0 > \sigma'$ were trapped, \emph{all} searches with $\sigma_0 < \sigma''$ were trapped, and \emph{some} searches with $\sigma'' \leq \sigma_0 \leq \sigma'$ were trapped. Figure~\ref{fig:trapstrength} illustrates this phenomenon for problem (C) by showing the fraction of successfully optimized searches as a function of the initial RFS. Also, the number of iterations required to reach an optimal solution grows as $\sigma_0$ decreases. The practical impact of the trap for each control problem is characterized by the values of $\sigma'$ and $\sigma''$ (which can be estimated from the data in Table~\ref{tab:rfsvar}), in comparison to the RFS values of optimal fields. 

For problems (A)--(C), the RFS $\sigma_{\text{opt}}$ of optimal fields is nearly invariant to the initial RFS $\sigma_0$. This result suggests that, for a given control problem, there exists a minimum RFS value $\sigma^*$ required to optimize the objective that is independent of the initial RFS $\sigma_0$. From the $\overline{\sigma_{\text{opt}}}$ data in Table~\ref{tab:rfsvar}, we estimate that $\sigma^* \approx 2.22 \times 10^{-2}$ for problem (A), $\sigma^* \approx 2.12 \times 10^{-1}$ for problem (B), and $\sigma^* \approx 5.44 \times 10^{-2}$ for problem (C). These values are physically reasonable, reflecting that the field strength must be sufficient to achieve at least one population flip in the control period in order to optimize successfully. That is, the Rabi frequency must satisfy $\Omega \gtrsim \pi / T$ or, equivalently, the RFS must satisfy $\sigma \gtrsim \pi/\omega T$ (where $\omega$ is a characteristic transition frequency). In addition, it is expected that $\omega T$ be at least a few multiples of $\pi$ in order to resolve the frequency of each resonant transition; if we assume that $\omega T \approx 10^1 \sim 10^2$, as is the case for OCT simulations in this paper, then we can estimate the minimum RFS for optimization as $\sigma^* \approx 10^{-1} \sim 10^{-2}$. For problems (A)--(C), examination of $\sigma_0$ values that result in attraction to the trap indicates that $\sigma'$, the field strength that corresponds to the upper attractive boundary of the trap, is smaller than $\sigma^*$ by at least an order of magnitude. This result indicates that searches starting with initial RFS values on the order of $\sigma^*$ will safely avoid search failure.

The effect of the trap is more noticeable for problems (D) and (E). As noted above, in these problems the final time $T$ is locked to a spectral transition of $H_0$ and fixed at $T = \pi/\alpha$, significantly constraining the control field. Several numerical studies \cite{MooreBrif2012, RivielloBrif-prep} indicate that such constraints can hamper OCT searches. Even for problems (D) and (E), however, $\sigma' \leq \sigma^*$. For every control problem studied, therefore, \textit{the region of the search space in which gradient searches sometimes fail is well separated from the region of the search space that contains optimal control fields}. Thus, optimizations that start with reasonable RFS values $\sim \sigma^*$ should avoid any traps.

While it is clear that the empirically measured values of $\sigma'$, $\sigma''$, and $\sigma^*$ vary between control problems, they also depend on algorithmic parameters. One can attempt to reduce search failure by demanding more accurate solutions of Eq.~\eqref{eq:searchalg}; in the case of \texttt{ode45}, this is accomplished by decreasing the absolute error tolerance $\tau$. However, greater solution accuracy generally increases the search effort required to optimize successfully.  An appropriately chosen error tolerance balances these demands and leads to searches of a practical effort with a minimal probability of search failure.  The \texttt{ode45} simulations summarized in Tables~\ref{tab:rfs1} and \ref{tab:rfsvar} use $\tau = 10^{-8}$. To examine how the error tolerance affects search failure, we performed additional optimizations of problem (A), varying the initial RFS over the range $10^{-5} \leq \sigma_0 \leq 10^{-1}$ and also varying the error tolerance over the range $10^{-12} \leq \tau \leq 10^{-2}$.  100 optimizations were performed for each value of $\sigma_0$ and $\tau$. 

\begin{table}[htbp]
\caption{\label{tab:tau}The number of failed optimizations for control problem (A). 100 optimization runs were performed for each value of $\sigma_0$ and $\tau$.}
\begin{ruledtabular}
\begin{tabular}{rcccccc}
\multicolumn{1}{c}{$\sigma_0$} & \multicolumn{6}{c}{Error tolerance $(\tau$)} \\
 & $10^{-2}$ & $10^{-4}$ & $10^{-6}$ & $10^{-8}$ & $10^{-10}$ & $10^{-12}$ \\ \hline
$10^{-1}$ & 0 & 0 & 0 & 0 & 0 & 0 \\ 
$5 \times 10^{-2}$ & 0 & 0 & 0 & 0 & 0 & 0 \\ 
$10^{-2}$ & 1 & 0 & 0 & 0 & 0 & 0 \\ 
$5 \times 10^{-3}$ & 47 & 0 & 0 & 0 & 0 & 0 \\ 
$10^{-3}$ & 100 & 90 & 0 & 0 & 0 & 0 \\
$5 \times 10^{-4}$ & 100 & 100 & 17 & 0 & 0 & 0 \\ 
$10^{-4}$ & 100 & 100 & 100 & 98 & 98 & 100 \\
$5 \times 10^{-5}$ & 100 & 100 & 100 & 100 & 100 & 100 \\ 
$10^{-5}$ & 100 & 100 & 100 & 100 & 100 & 100 \\
\end{tabular}
\end{ruledtabular}
\end{table}
Table~\ref{tab:tau} shows that values of $\tau > 10^{-8}$ lead to search failure at larger values of $\sigma_0$.  However, decreasing $\tau$ below $10^{-8}$ does not substantially decrease the initial RFS at which searches fail. This result suggests that the value $\sigma_0 = 5 \times 10^{-4}$ is the smallest at which the \texttt{ode45} algorithm can avoid search failure entirely for problem (A).  It also confirms that $\tau = 10^{-8}$ is an appropriate choice of error tolerance for this control problem, since smaller values do not reduce the likelihood of failure. However, it is possible that other gradient methods could optimize successfully using even smaller $\sigma_0$ values.

Many factors affect whether gradient searches fail for control problems whose landscapes contain a second-order trap. The minimum initial RFS required to avoid search failure, $\sigma'$, is both problem- and algorithm-dependent, and gradient optimizations of problems (A)--(E) may fail at very low initial RFS even when an accurate algorithm is employed. However, such weak initial fields are not likely to be encountered in OCE or OCT because the RFS required for optimal fields, $\sigma^*$, is found to be significantly larger than $\sigma'$ for well-designed searches. Therefore, it is extremely unlikely that gradient searches starting with physically reasonable RFS values of $\sim \sigma^*$ will fail due to the presence of second-order singular traps on the control landscape. This result, along with the artificial and physically unmotivated relationship between the system Hamiltonian and objective imposed by all control problems with a second-order trap, leads us to conclude that the potential presence of such traps on the landscape is of negligible importance for practical optimal control experiments and simulations.

\subsection{Singular critical points lacking Hessian analysis}
\label{sec:saddles}

Although the singular critical points in problems (A)--(E) are second-order traps, this is not necessarily true of all singular critical points that violate condition (2). One control problem described in \cite{PechenTannor2011} contains a singular critical point for which Hessian analysis was not performed:

\begin{enumerate}
\item[(F)] Maximize $J_{\theta}$ for a closed $N$-level system. Define the states
\begin{equation*}
| \psi_{\pm} \rangle = \frac{|i \rangle \pm e^{i \chi}|j \rangle}{\sqrt{2}},
\end{equation*}
where $|i \rangle$ and $|j \rangle$ are distinct eigenstates of $H_0$. Similarly, define the states $| \psi'_{\pm} \rangle$ by employing the same choice of $|i \rangle$ and $|j \rangle$ but a different phase $\chi' \neq \chi$.  Suppose that $\rho_0 = | \psi_+ \rangle \langle \psi_+ |$ and that $Q$ is an arbitrary Hermitian operator with $| \psi_+ \rangle$ in its null space. If $\langle i | \mu | i \rangle = \langle j | \mu | j \rangle$, the landscape $J_{\theta}[\varepsilon(t)]$ corresponding to the observable
\begin{equation*}
\theta = e^{-i H_0 T/\hbar} (| \psi'_+ \rangle \langle \psi'_+ | + Q) e^{i H_0 T/\hbar} ,
\end{equation*}
has a singular critical point at $\varepsilon(t) = 0$ \cite{PechenTannor2011}. Unlike the examples in Sec.~\ref{sec:traps}, there is no Hessian analysis indicating that this critical point is a trap. In our simulations, we used a three-level system:
\begin{equation*}
H_0 = \begin{pmatrix} 0 & 0 & 0 \\ 0 & 10 & 0 \\ 0 & 0 & 30 \end{pmatrix}, \ 
\mu = \begin{pmatrix} 0 & 1 & 0.5 \\ 1 & 0 & 1 \\ 0.5 & 1 & 0 \end{pmatrix},
\end{equation*}
and defined the Hermitian operator $Q$ as
\begin{equation*}
Q = \sum_{k,l \neq i,j} q_{kl} |k \rangle \langle l | 
+ | \psi_- \rangle \langle \psi_- | .
\end{equation*}
We performed a numerical study on problem (F) similar to those performed on problems (A) - (E). For each simulation, we randomly selected the eigenstates $| i \rangle$ and $| j \rangle$, phases $\chi, \chi' \in [0,2 \pi )$, and coefficients $q_{kl} \in [0,1]$. The final time was $T = 8$, and the control period was divided into $L = 255$ intervals.
\end{enumerate}

\begin{table}[htbp]
\caption{\label{tab:rfsdyn}Optimization results for control problem (F), with initial fields of different RFS. 100 optimization runs were performed for each value of $\sigma_0$.}
\begin{ruledtabular}
\begin{tabular}{rrr}
$\sigma_0$ & \# failed & MSE \\ \hline
$5 \times 10^{-5}$ & 0 & 70 \\
$10^{-5}$ & 0 & 76 \\
$5 \times 10^{-6}$ & 0 & 73 \\
$10^{-6}$ & 0 & 66 \\
$5 \times 10^{-7}$ & 0 & 77 \\
\end{tabular}
\end{ruledtabular}
\end{table}
We performed 100 optimization runs for each $\sigma_0$ value over the range $5 \times 10^{-5} \geq \sigma_0 \geq 5 \times 10^{-7}$, in order to determine whether the control landscape for problem (F) has a trapping region analogous to those observed for problems (A)--(E). As reported in Table~\ref{tab:rfsdyn}, we found that all search trajectories with $\sigma_0 \geq 5 \times 10^{-7}$ optimized successfully.  Fields at this RFS are within the attractive regions surrounding the traps in problems (A)--(E), suggesting that the singular critical point at zero field in problem (F) is not a trap. The numerically evaluated Hessian matrix $\mathsf{H}(t,t')$ at $\varepsilon(t) = 0$ is indefinite, confirming that the critical point is a saddle. These results are consistent with an earlier study \cite{WuLongDominy2012}, which used a special algorithm to find singular critical points for 100 control problems with randomly selected Hamiltonians. All of the singular critical points found by this method corresponded to non-constant control fields; they exhibited no trapping effect on gradient searches and were numerically confirmed to be saddles through Hessian analysis, similar to what we observe here for the zero-field singular critical point in problem (F). In the context of this prior study, our results here support the conclusion that almost all singular critical points that violate condition (2) do not impede gradient optimization.

\subsection{Absence of singular critical points}

For some control problems, there is no proof of whether or not the landscape contains singular critical points. One such problem was defined in \cite{FouquieresSchirmer2010}: 

\begin{enumerate}
\item[(G)] Minimize $\tilde{J}_W$ (see Eq.~\eqref{eq:w2}) for an eight-level system consisting of three Ising-coupled qubits:
\begin{equation*}
\begin{split}
&2 H(t) = Z_1 Z_2 + Z_2 Z_3 + \varepsilon_1(t) X_1 + \varepsilon_2(t) Y_1 \\
&+ \varepsilon_3(t) X_2 + \varepsilon_4(t) Y_2 + \varepsilon_5(t) X_3 + \varepsilon_6(t) Y_3 ,
\end{split}
\end{equation*}
where $X_1 = \sigma_x \otimes I \otimes I$, $Y_2 = I \otimes \sigma_y \otimes I$, etc. There are six control fields. The target unitary transformation is a quantum Fourier transform (QFT) gate: 
\begin{equation*}
W = \sum_{j,k=1}^8 \frac{\exp(2 \pi i (m + \frac{1}{4})/8)}{\sqrt{8}} \xi^{jk} | j \rangle \langle k | ,
\end{equation*}
where $\xi = \exp (-2 \pi i/8)$ and $m$ is an integer.
\end{enumerate}

In \cite{FouquieresSchirmer2010}, 1000 OCT optimizations of problem (G) were performed using a control period $T = 8$ divided into $L = 140$ intervals, and a small number of them became trapped at suboptimal fidelities. In evolution-operator control, shortening the control period may be an attractive feature in some circumstances \cite{NielsenChuang2000}; however, theoretical analysis and numerical simulations have both shown that a sufficiently large control time $T$ is necessary in order to reach a global optimum. Time-optimal control, the problem of finding control fields that achieve a desired fidelity in the minimal time necessary, addresses this issue and has been explored in both OCT and OCE \cite{Khaneja2001, KhanejaHeitmann2007, MooreBrif2012}. In this work, we performed OCT simulations for problem (G) with $T = 6,7,8,9,10$ and $L = 140$; 200 optimizations were performed for each value of $T$ for $T \geq 7$ and 25 optimizations were performed for $T = 6$.

\begin{figure}[htbp]
\centering
\includegraphics[width=8cm]{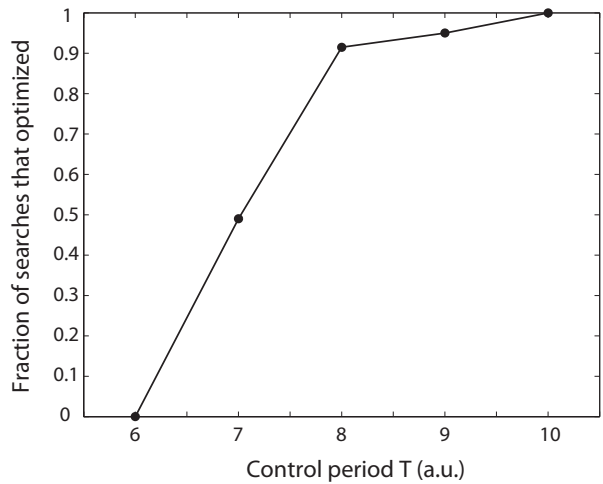}
\caption{The fraction of searches for control problem (G) that optimized successfully, as a function of the control period $T$. 200 optimization runs were performed for each value of $T$ for $T \geq 7$ and 25 optimizations were performed for $T = 6$.}
\label{fig:schirmerruns}
\end{figure}
All optimization runs succeeded when the control period was at least $T = 10$, as illustrated in Figure~\ref{fig:schirmerruns}. This result suggests that the failed optimization runs \cite{FouquieresSchirmer2010} for $T \leq 9$ are caused by violation of condition (3) (constraining the control field) rather than violation of condition (2) (a rank-deficient Jacobian). Problems with significantly constrained control resources (such as a limited control time $T$ in problem (G)) can arise, and therefore significant attention should be paid in practice to alleviate such constraints. This common situation is fundamentally different from the circumstances surrounding problems (A)--(F), where singular critical points violating condition (2) are present on the landscape.

\section{Conclusions}
\label{sec:concl}

The success of quantum control experiments and simulations has prompted several works devoted to theoretical analysis of the landscape critical topology \cite{HoRabitz2006JPPA, WuRabitzHsieh2008JPA, Altafini2009, ChakrabartiRabitz2007review, WuPechenRabitz2008JMP, RabitzHsiehRosenthal2004, RabitzHoHsieh2006PRA, RabitzHsiehRosenthal2005PRA, DominyRabitz2008JPA, HsiehHoRabitz2008CP, ShenHsiehRabitz2006JCP}. Collectively, these studies contend that the absence of local optima on the control landscape is responsible for the favorable results in OCE and OCT. This trap-free structure depends upon three conditions:  controllability, a full rank Jacobian matrix $\delta U_T / \delta \varepsilon(t)$, and an unconstrained control field $\varepsilon(t)$. This paper has investigated how gradient searches are affected by violation of the second condition.

The landscapes for optimal control problems may contain singular critical points, at which the Jacobian is rank deficient. However, empirical evidence suggests that the conditions necessary to yield singular critical points are sufficiently strict that regular critical points are far more common \cite{WuLongDominy2012}. Moreover, only a very small subset of singular critical points have been shown to prevent gradient searches from optimizing, and every singular critical point found to cause search failure has been identified analytically as a second-order trap. These traps correspond to constant fields, particularly $\varepsilon(t) = 0$; such controls have negligible physical relevance and have not been observed to hinder any quantum control experiments. Numerical and analytical evidence from the quantum control literature indicates that second-order traps that violate condition (2) rarely impede a gradient search for optimal controls. All known traps have only been observed when the control system and objective satisfy sets of unusually stringent conditions as described in Sec.~\ref{sec:traps}.  Any deviation from these artificially imposed conditions eliminates the suboptimal trap, and the evidence shows that the great majority of control landscapes lack any traps. We have also shown that the failure of numerical optimization for some control problems \cite{FouquieresSchirmer2010} is due not to a rank-deficient Jacobian, but rather to a significant constraint on the control field \cite{MooreBrif2012, RivielloBrif-prep}.

In this work, thousands of numerical simulations have been performed on control problems whose landscapes contain a second-order trap at zero field. They showed that gradient searches only fail when the strength of the initial control field is much below what is required for optimal fields, i.e., when the initial field is very close to the trap. Such weak fields correspond to an infinitesimally small region of the control space, and searches originating outside of this region always optimized successfully. Although no second-order traps have been identified at non-constant control fields, it is reasonable to hypothesize that, if they were to exist, their effect on gradient optimization would similarly be limited to searches that begin very close to them. Our studies in this work have therefore led us to conclude that gradient searches performed on controllable systems are very unlikely to fail, even when the Jacobian rank condition (one of the requirements for a trap-free landscape) is violated. This result, combined with the stringent conditions required to observe second-order traps, effectively eliminates singular critical points as significant obstacles to gradient-based quantum optimal control. 

\acknowledgments

The authors acknowledge support from the Department of Energy under grant DE-FG02-02ER15344 and the Army Research Office under grant W911NF-13-1-0237. RBW acknowledge support from NSFC under Grant Nos. 61374091, 60904034 and 61134008. This work was supported by the Laboratory Directed Research and Development program at Sandia National Laboratories. Sandia National Laboratories is a multi-program laboratory managed and operated by Sandia Corporation, a wholly owned subsidiary of Lockheed Martin Corporation, for the U.S. Department of Energy's National Nuclear Security Administration under contract DE-AC04-94AL85000.

\raggedright
\bibliographystyle{apsrev4-1}
\bibliography{riviellorabitz_singularpoints2}

\end{document}